\documentclass[conference,a4paper]{APSIPA2025}
\usepackage{amsmath}
\usepackage{graphicx}
\usepackage{multirow}
\usepackage{threeparttable}
\usepackage[backend=biber,style=ieee,]{biblatex}
\usepackage{subcaption} 
\usepackage{float}
\usepackage{stfloats}
\usepackage{booktabs}
\usepackage{multicol}
\usepackage{caption}
\usepackage{url}
\usepackage{biblatex}
\usepackage{amsmath,amssymb,amsfonts}

\usepackage{geometry}
\usepackage{fancyhdr}

\fancypagestyle{firststyle}{
  \fancyhf{}
  \fancyhead[C]{2025 Asia Pacific Signal and Information Processing Association Annual Summit and Conference (APSIPA ASC)}
}

\begin{document}
\setlength{\dbltextfloatsep}{12pt} 

\title{Emotional Text-To-Speech Based on Mutual-Information-Guided Emotion-Timbre Disentanglement}

\author{
\authorblockN{
Jianing Yang\authorrefmark{1}\authorrefmark{3},
Sheng Li\authorrefmark{2},
Takahiro Shinozaki\authorrefmark{2},
Yuki Saito\authorrefmark{1},
Hiroshi Saruwatari\authorrefmark{1}
}

\authorblockA{
\authorrefmark{1}The University of Tokyo, Japan, \authorrefmark{2}Institute of Science Tokyo, Japan.}


\authorblockA{
\authorrefmark{3}
E-mail: baleyang@g.ecc.u-tokyo.ac.jp\\}
}

\maketitle
\thispagestyle{firststyle}
\pagestyle{fancy}
\cfoot{}

\begin{abstract}
  Current emotional Text-To-Speech (TTS) and style transfer methods rely on reference encoders to control global style or emotion vectors, but do not capture nuanced acoustic details of the reference speech. To this end, we propose a novel emotional TTS method that enables fine-grained phoneme-level emotion embedding prediction while disentangling intrinsic attributes of the reference speech. The proposed method employs a style disentanglement method to guide two feature extractors, reducing mutual information between timbre and emotion features, and effectively separating distinct style components from the reference speech. Experimental results demonstrate that our method outperforms baseline TTS systems in generating natural and emotionally rich speech. This work highlights the potential of disentangled and fine-grained representations in advancing the quality and flexibility of emotional TTS systems.\footnote{The synthesized audio samples are available at \url{https://baleyang.github.io/emotion-timbre-disentangled-tts/}}
\end{abstract}

\section{Introduction}

\begin{figure*}[t]
    \centering
    \includegraphics[width=\textwidth]{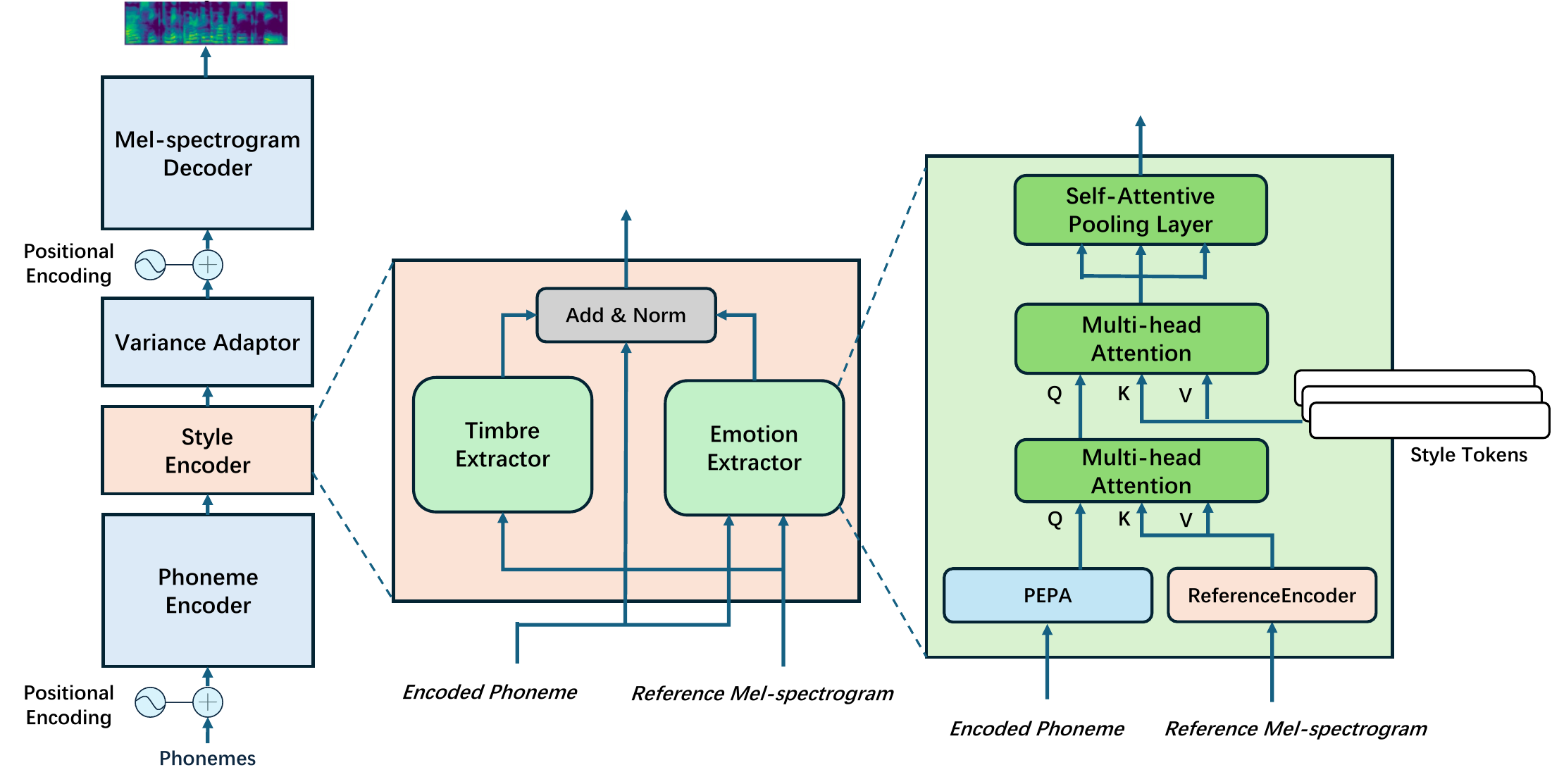}
    \caption{\normalfont Overview of the proposed TTS model and its main modules. \textbf{Left}: The end-to-end TTS pipeline adopts the FastSpeech 2 backbone; a Style Encoder is inserted after the Phoneme Encoder, followed by the Variance Adaptor and the Mel-spectrogram Decoder. \textbf{Center}: Architecture of the Style Encoder, in which separate Timbre and Emotion Extractor are combined through a residual Add \& Norm operation. \textbf{Right}: Detailed design of the Emotion Extractor, generates phoneme-level emotion embeddings.}
    \label{fig:overall}
\end{figure*}

Deep learning has significantly advanced Text-To-Speech (TTS) technology, surpassing early statistical models that struggled with naturalness and expressiveness. The introduction of deep neural networks (DNNs)~\cite{Vaswani2017AttentionIA} as an acoustic model improved speech fidelity and intelligibility by capturing complex relationships between input text and output speech in TTS. Autoregressive generative models~\cite{Tacotron2, Oord2016WaveNetAG, TransformerTTS} enabled end-to-end TTS, which represents the mapping from input text to output speech by stacked DNN modules and enhances synthesis quality and robustness toward prosody drift and mis-alignment in long-form synthesis. More recently, non-autoregressive models~\cite{peng2020non, ren2019fastspeech, Ren2020FastSpeech2F} have attracted increasing attention. In contrast to autoregressive models, which generate speech frames sequentially, non-autoregressive models predict all output frames in parallel. This parallelization significantly speeds up inference, greatly enhances the efficiency of TTS. 


Despite significant advances in naturalness and diversity of synthetic speech by TTS, achieving precise and expressive emotional TTS remains a challenging task. Emotional TTS, particularly in zero-shot settings, aims to generate speech that matches the timbre, emotion, and prosody of a few seconds of reference speech. Traditional methods typically encode the reference speech into a global style vector~\cite{GSTs, StyleSpeech, SkerryRyan2018TowardsEP}, which is then fused with the output of a phoneme encoder. Although these approaches effectively capture the overall style, they often struggle to model the phoneme-level variation in emotion and prosody. Moreover, compressing the reference speech into a single global embedding risks losing crucial prosodic details, limiting the expressiveness and control over the synthesized speech.

To address these limitations, we propose a novel emotional TTS method that (i) predicts fine-grained, phoneme-level emotion embeddings, and (ii) disentangles those embeddings from global timbre information through mutual-information minimization. Central to our method is a dedicated Style Encoder, which comprises two parallel extractors: a global Timbre Extractor, and a phoneme-aware Emotion Extractor that aligns reference acoustics with target phonemes to produce an emotion embedding sequence. An unsupervised Mutual Information Neural Estimation (MINE)~\cite{MINE} explicitly pushes the two representations apart, ensuring that the timbre embedding remains speaker-specific information, while the emotion embeddings capture only prosodic nuance, allowing the model to synthesize speech that is simultaneously timbre-consistent and emotionally expressive.

Experiments demonstrate that our method outperforms strong baselines, including Global Style Token(GST)~\cite{GSTs}, StyleSpeech~\cite{StyleSpeech}, MIST~\cite{Hu2020UnsupervisedSA}, and DC Comix TTS~\cite{dccomixtts}, on both subjective and objective metrics. t-SNE visualizations further reveal well-separated emotion clusters, confirming effective disentanglement. Our results highlight the value of combining phoneme-level emotion modeling with principled feature disentanglement for expressive, high-fidelity emotional TTS.

\section{Related Works}
TTS has recently progressed from sentence‑level style transfer to fine‑grained prosody control, yet three technical lines still dominate the literature. Below we summarize each line and highlight the open problem our study tackles.

\paragraph{\textbf{Global \& hierarchical emotion embeddings}}
~\cite{GSTs} first proposed GST---a bank of learnable tokens attended by a reference encoder---to condense an utterance into a single ``style vector.'' Subsequent works refined this idea: StyleSpeech~\cite{StyleSpeech} injects the vector into every encoder/decoder block via Style‑Adaptive LayerNorm (SALN); Tacotron‑GST~\cite{SkerryRyan2018TowardsEP} piles GSTs hierarchically to capture speaking styles spanning words to paragraphs. Although these systems improve expressiveness, their utterance‑level embeddings cannot localize phoneme‑wise variations and therefore provide only coarse control over emotion and rhythm.

\paragraph{\textbf{Prosody modeling via neural codecs}}
Discrete token representations have recently become a focal point in speech modeling research. Neural audio codecs based on vector‑quantized variational autoencoders—most notably SoundStream~\cite{SoundStream} and EnCodec~\cite{EnCodec}—transform continuous waveforms into sequences of codebook indices, yielding a compact ``speech language'' that lightweight sequence‑to‑sequence decoders can handle efficiently. Building on this foundation, several studies have investigated how such tokens can capture prosody for zero‑shot TTS. A representative example is DC Comix TTS~\cite{dccomixtts}, which tokenizes a reference signal with an EnCodec‑style front end and conditions its decoder on a style embedding derived from the resulting code sequence, achieving high‑fidelity speech from unseen speakers. However, fixed‑rate quantization still blurs micro‑prosodic cues—such as subtle emotional nuances and pitch inflections—and the reliance on a global utterance‑level embedding limits phoneme‑level expressive control.

\paragraph{\textbf{Feature disentanglement}}
Most disentanglement studies focus on separating content from a holistic ``style'' embedding. Typical methods include auxiliary classifiers~\cite{FHVAE}, adversarial objectives~\cite{9362098}, and mutual information minimization (MIST;~\cite{Hu2020UnsupervisedSA}; ProsodySpeech;~\cite{ProsodySpeech}). Although effective in maintaining intact linguistic information, \textbf{they overlook the inter‑style entanglement}---timbre, emotion, and local prosody still co‑exist in the same vector, making controllable synthesis difficult and hurting zero‑shot generalization. 


\section{Model Architecture}
To address the limitations of global style embeddings and leverage MINE for effective speech feature disentanglement, we propose a novel model architecture for emotional TTS that integrates MINE into the style encoding process. Our model introduces two key innovations:

\noindent\textbf{1. Phoneme-level emotion embedding prediction}: The model predicts emotional information at the phoneme level while treating timbre as a global feature, enabling TTS that closely matches the style of the reference speech.

\noindent\textbf{2. Effective disentanglement of speech features}: By guiding the feature extractor to focus on distinct attributes of the reference mel-spectrogram, we successfully decouple timbre and emotional features, improving the style controllability of synthetic speech. 

By combining global timbre prediction with fine-grained emotional modeling and employing MINE for feature disentanglement, our method overcomes the limitations of existing approaches. This improvement significantly enhances the style similarity and expressiveness of the synthetic speech.

\subsection{Model Backbone}
The overall architecture of the proposed model is illustrated in Fig.~\ref{fig:overall}. It is based on FastSpeech 2(FS2)~\cite{Ren2020FastSpeech2F}, consisting of Encoder-Decoder networks with Variance Adaptor, which following the original FS2 method. To enable emotional TTS, we introduce a Style Encoder after the Phoneme Encoder to predict style-specific representations.

Specifically, the Phoneme Encoder consists of four Feed-Forward Transformer (FFT) blocks, while the mel-spectrogram Decoder includes six FFT blocks. The Variance Adaptor comprises a length regulator, pitch predictor, and energy predictor, each implemented using two 1D convolution layers with 256 filters. The structure and design of the Style Encoder are detailed in the next section.

\subsection{Style Encoder}
\label{sec:style_encoder}


The architecture of our Style Encoder is composed of two parallel modules---a \textbf{Timbre Extractor} and an \textbf{Emotion Extractor}---that independently leverage the Phoneme Encoder output and a reference mel-spectrogram to derive phoneme-level style information. The central assumption is that timbre remains relatively stable and invariant to specific textual content, whereas emotion and prosody vary significantly with different inputs.

We adopt GST-based method~\cite{GSTs} for timbre extraction. Specifically, a Reference Encoder first processes the reference mel-spectrogram to produce an intermediate representation, which is then passed through a style token layer to obtain a global timbre embedding, $\mathbf{F}_{\text{timbre}}$. 

For Emotion Extractor, we design a phoneme-aware architecture to capture fine-grained emotional nuances. We begin by encoding the reference mel-spectrogram using the same Reference Encoder employed in the Timbre Extractor. 

To bridge the representational gap between the Phoneme Encoder and the Emotion Extractor, we introduce a lightweight Phoneme-Emotion Projection Adapter (PEPA). Implemented as two successive 1-D convolutional layers, PEPA projects the phoneme embeddings generated by the phoneme encoder that is pre-trained exclusively on neutral speech in stage one and kept fixed in stage two (as described in Section~\ref{sec:implementation})—into the prosody-rich acoustic space learned by the Reference Encoder. This projection supplies each phoneme with temporally aligned emotional context, thereby mitigating the mismatch caused by the two-stage training scheme and enabling fine-grained fusion of linguistic and emotional cues.

We then invoke a multi-head cross-attention module that treats the projected phoneme embeddings as queries while using the reference-encoder emotion features as keys and values. For each phoneme position $i$, the attention matrix yields weights $\alpha_{ij}$ over the emotion sequence $e_j$; the resulting representation is defined as the weighted sum of $\alpha_{ij}$ and $e_j$:
$$
\tilde{p}_i=\sum_{j}\alpha_{ij}\,e_j .
$$
This operation endows every phoneme with a custom blend of emotional cues proportional to its affinity with each reference frame, producing phoneme-synchronous emotional features that seamlessly fuse linguistic and affective information. Notably, we do \emph{not} include positional encoding for the reference mel-spectrogram features, preventing potential ``content leakage'' and ensuring robust TTS when the reference mel-spectrogram and target text are mismatched. Following this alignment, we employ another multi-head cross-attention mechanism combined with a style token layer to generate the emotion embedding sequence $\mathbf{F}_{\text{emotion}}$. We then apply a Self-Attentive Pooling Layer~\cite{Chen2022SelfAttentivePF} to smooth transitions between adjacent phonemes. The final output of these processes is the emotion embedding, $\mathbf{F}_{\text{emotion}\_\text{smooth}}$.

Finally, the Phoneme Encoder output is combined with $\mathbf{F}_{\text{emotion}\_\text{smooth}}$ via element-wise addition and then broadcast-summed with $\mathbf{F}_{\text{timbre}}$. After applying layer normalization, the Style Encoder produces its final representation, effectively capturing both global timbre and fine-grained emotional cues. 

\subsection{Style Disentanglement}

\begin{figure}[t]
    \centering
    \includegraphics[width=0.47\textwidth]{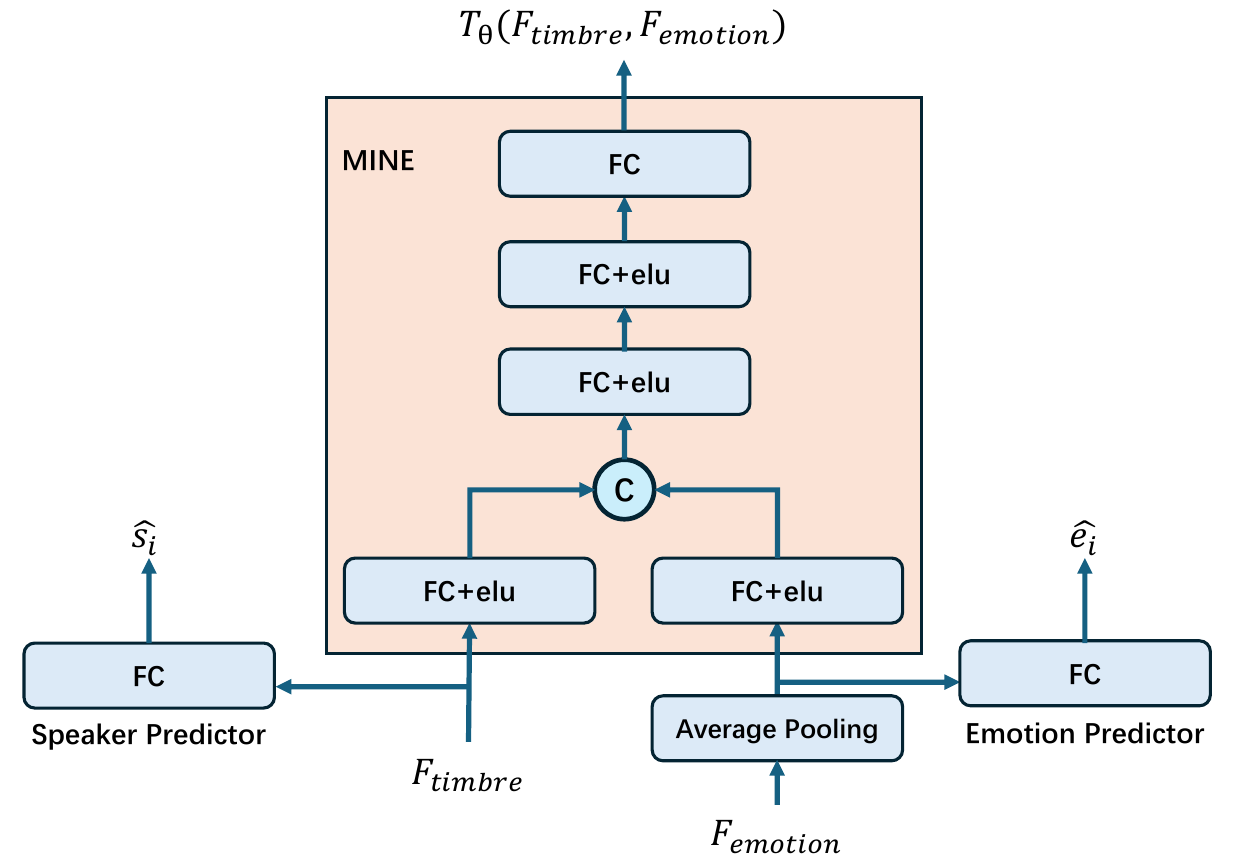}
    \caption{\normalfont Proposed architecture for timber-emotion disentanglement based on mutual information minimization}
    \label{StyleDisentanglement}
\end{figure}

Given the dual-feature extractors (Timbre Extractor and Emotion Extractor) in our model, it is essential to disentangle their respective outputs to ensure that they capture distinct speech attributes. To achieve this, we employ MINE to disentangle the outputs of these two extractors.

Previous studies have attempted to achieve feature disentanglement by minimizing the mutual information between style embeddings. However, these methods face a common issue: the lack of explicit guidance for the disentanglement process. Simply relying on MINE to minimize the mutual information between style embeddings leaves the model with no clear optimization direction, hindering its ability to effectively separate features. To address this, as shown in Fig.~\ref{StyleDisentanglement}, we not only minimize the mutual information between the emotion embedding $\mathbf{F}_{\text{emotion}}$ and the timbre embedding $\mathbf{F}_{\text{timbre}}$ using MINE, but also guide the disentanglement by explicitly predicting emotion and speaker labels from $\mathbf{F}_{\text{emotion}}$ and $\mathbf{F}_{\text{timbre}}$, respectively. This approach provides clear optimization objectives, enabling effective emotional speech synthesis.

Specifically, $\mathbf{F}_{\text{emotion}}$ is a sequence of phoneme-level emotion embeddings. We apply average pooling to aggregate this sequence into a global emotion embedding $\mathbf{F}_{\text{emotion}\_\text{global}}$. Then, fully connected (FC) layers, named the \textit{Emotion Predictor} and \textit{Epeaker Predictor}, are used to predict the corresponding emotion label and speaker label, providing explicit objectives for optimizing the style extractors.

To estimate and suppress the mutual information between the global-emotion embedding $\mathbf{F}_{\text{emotion}\_\text{global}}$ and the timbre embedding $\mathbf{F}_{\text{timbre}}$, we follow the Donsker–Varadhan (DV) variational formulation of the KL divergence \cite{Donsker1975AsymptoticEO} that underpins recent works \cite{Hu2020UnsupervisedSA, ProsodySpeech, paul21_interspeech, 9362098}. For any measurable scoring function $T\colon (Y,Z)\!\mapsto\!\mathbb{R}$, the DV inequality gives
$$
\mathcal{I}(Y,Z)\;\ge\;\hat{\mathcal{I}}_T(Y,Z)
\;=\;
\mathbb{E}_{P_{Y,Z}}[T]\;-\;\log\!\bigl(\mathbb{E}_{P_Y\!\otimes P_Z}[e^{T}]\bigr).
$$
MINE~\cite{MINE} instantiates $T$ as a trainable neural network $T_\theta$ and maximizes $\hat{\mathcal{I}}_{T_\theta}$ with respect to the model parameter $\theta$, thereby tightening the lower bound.

In our implementation, $T_\theta$ first processes $\mathbf{F}_{\text{emotion}\_\text{global}}$ and $\mathbf{F}_{\text{timbre}}$ through two independent fully-connected (FC) layers, each followed by an ELU activation \cite{Clevert2015FastAA}. The resulting vectors are then concatenated and passed to a three-layer FC head whose first two layers again use ELU, while the final layer outputs a scalar score.

During training, the mutual information estimator is optimized by maximizing the negative mutual information $-\hat{\mathcal{I}}_{T_\theta}(Y, Z)$ to capture the dependency between $\mathbf{F}_{\text{emotion}\_\text{global}}$ and $\mathbf{F}_{\text{timbre}}$. Simultaneously, the Timbre Extractor and Emotion extractor are updated by minimizing the mutual information $\hat{\mathcal{I}}_{T_\theta}(Y, Z)$, effectively reducing the overlap between the two embeddings. This process ensures that the emotional and timbre features are disentangled, achieving robust style disentanglement.

By combining mutual information minimization with explicit supervision via emotion and speaker labels, our method overcomes the challenges faced by previous methods, enabling the extraction of distinct and independent style attributes for more expressive and controllable emotional TTS.

\section{Implementation}
\label{sec:implementation}
As indicated in~\cite{Hu2020UnsupervisedSA, ProsodySpeech, paul21_interspeech}, obtaining a ``clean'' encoder whose output phoneme representations do not carry emotion-related information is critical. In the first stage of our method, we therefore train the FastSpeech 2 model \emph{without} the Style Encoder and use only speech samples labeled with the \emph{neutral} emotion category to avoid style-induced variability. The objective in this stage is given by:
\begin{align*}
\label{eq:stage1}
\mathcal{L}_1 &= \mathcal{L}_{\mathrm{recons}} + \lambda_1\,\mathcal{L}_{\mathrm{dur}},
\end{align*}
where $\mathcal{L}_{\mathrm{recons}}$ denotes the reconstruction loss on the predicted mel-spectrogram, and $\mathcal{L}_{\mathrm{dur}}$ denotes the duration prediction loss. We set $\lambda_1 = 1.0$ throughout our experiments. Once this first-stage training converges, we obtain an encoder that is less sensitive to emotion-related information.

In the second stage, we incorporate the Style Encoder into the model and initialize all encoder parameters using the pre-trained weights from stage one (and freeze these parameters). To ensure the Timbre Extractor and the Emotion Extractor focus on distinct aspects of the reference mel-spectrogram, we adopt a mutual information minimization scheme based on MINE. Specifically, we alternate between updating the TTS model and the MI estimator to encourage the Timbre Extractor and the Emotion Extractor to capture non-overlapping information. We augment our loss with additional pitch and energy terms, as well as classification losses for speaker and emotion. The second-stage objective is defined as:
\begin{align*}
\mathcal{L}_{2} \;=\;& \mathcal{L}_{\mathrm{recons}} 
\;+\; \lambda_1 \mathcal{L}_{\mathrm{dur}} 
\;+\; \lambda_2 \mathcal{L}_{\mathrm{pitch}} \nonumber\\
& +\; \lambda_3 \mathcal{L}_{\mathrm{energy}}  
\;+\; \lambda_4 \mathcal{L}_{\mathrm{emotion}}
\;+\; \lambda_5 \mathcal{L}_{\mathrm{speaker}}\nonumber\\
& +\; \lambda_6 
    \mathrm{ReLU}\bigl(\hat{\mathcal{I}}_{T_\theta}\bigl(\mathbf{F}_{\mathrm{timbre}}, \mathbf{F}_{\mathrm{emotion\_global}}\bigr)\bigr),
\end{align*}
where $\mathcal{L}_{\mathrm{pitch}}$ and $\mathcal{L}_{\mathrm{energy}}$ measure prediction errors for pitch and energy, respectively, and $\mathcal{L}_{\mathrm{emotion}}$ and $\mathcal{L}_{\mathrm{speaker}}$ are cross-entropy classification losses for emotion and speaker identification, respectively. The term $\hat{\mathcal{I}}_{T_\theta}(\mathbf{F}_{\mathrm{timbre}}, \mathbf{F}_{\mathrm{emotion\_global}})$ is the estimated mutual information, and $-\hat{\mathcal{I}}_{T_\theta}$ is optimized in the MI estimator to capture any correlation between timbre and emotion features. We empirically set $\lambda_1 = 1.0$, $\lambda_2 = 1.0$, $\lambda_3 = 1.0$, $\lambda_4 = 1.0$, $\lambda_5 = 1.0$, $\lambda_6 = 0.1$, throughout our experiments. By alternating gradient updates between the TTS model and the MI estimator, we encourage the Timbre Extractor and Emotion Extractor to learn disentangled representations.

\section{Experiments}
\subsection{Dataset}
We used emotional speech databases as the main source for our experiments. Specifically, we used the ESD (Emotional Speech Dataset)~\cite{Zhou2021EmotionalVC}, which contains 350 parallel utterances spoken by 10 native English speakers and 10 native Chinese speakers, covering five distinct emotion categories: neutral, happy, angry, sad, and surprise. In our experiments, we selected the English subset of the dataset. The data was randomly split into 80\% for training, 10\% for validation, and 10\% for testing.

\subsection{Baselines}
We conducted a comparison between our model and various baselines.



\noindent\textbf{FS2 + GST.} An expressive TTS model that integrates GSTs~\cite{GSTs} into the FS2 method to capture diverse speaking styles.

\noindent\textbf{StyleSpeech~\cite{StyleSpeech}.} An expressive TTS model employing a reference encoder to produce a style embedding, which in turn modulates the output of SALN layers via gain and bias parameters.

\noindent\textbf{FS2 + MIST~\cite{Hu2020UnsupervisedSA}.} A FS2-based model introducing MIST, which reduces the mutual information between the phoneme encoder and the style embedding. This encourages disentanglement of style and content for improved expressive synthesis.

\noindent\textbf{DC Comix TTS~\cite{dccomixtts}.} A variant that replaces GST with a reference encoder based on discrete code.

\subsection{Implementation Details}
We systematically compared our proposed method with baseline methods, each built on the same FS2 architecture. For a fair comparison, all models followed the same DNN architectures for the encoder, decoder, and variance adaptors. We used the Adam optimizer with $\beta$ = (0.9, 0.98), $\epsilon = 1 \times 10^{-8}$, and the learning rate $\ell_t$ follows~\cite{Vaswani2017AttentionIA}. The mini-batches contained 64 samples, and all models were trained for 60 k optimization steps. For vocoder, We initially employ the official \texttt{UNIVERSAL\_V1}\footnote{\url{https://github.com/jik876/hifi-gan}} version of the pre-trained HiFi-GAN~\cite{Kong2020HiFiGANGA}, and subsequently perform fine-tuning on the ESD training set to better adapt the vocoder to our data. This adapted vocoder was then used to synthesize the final speech waveforms from the generated mel-spectrograms.

\subsection{Evaluation Metrics}


\begin{table*}[t]
\centering
\caption{\normalfont Evaluation results: MOS, MCD, and UAA}
\begin{tabular}{@{}lccccc@{}}
\toprule
\textbf{Model} & \textbf{MOS(↑)} & \textbf{SMOS(↑)}     & \textbf{MCD(↓)}    & \textbf{UAA(↑)}  \\ \midrule
FS2 + GST     & $3.44\pm0.11$    & $3.12\pm0.07$        & $8.75\pm0.22$     & 74.22\%    \\
StyleSpeech     & $2.95\pm0.11$   & $3.26\pm0.06$         & $9.09\pm0.23$      & 82.22\%    \\
FS2 + MIST    &  $3.62\pm0.10$    & $2.89\pm0.08$          & $8.65\pm0.22$      & 76.17\%  \\ 
DC Comix TTS    &  $3.59\pm0.11$   & $2.97\pm0.08$         & $9.01\pm0.24$      & 58.79 \%  \\ \midrule
\textbf{Proposed}  & $\mathbf{3.63\pm0.10}$  & $\mathbf{3.41\pm0.06}$      & $\mathbf{8.23\pm0.22}$ & $\mathbf{82.42\%}$\\ \bottomrule
\end{tabular}
\label{tab:results1}
\end{table*}

\begin{figure*}[t]
    \centering
    \includegraphics[width=0.9\textwidth]{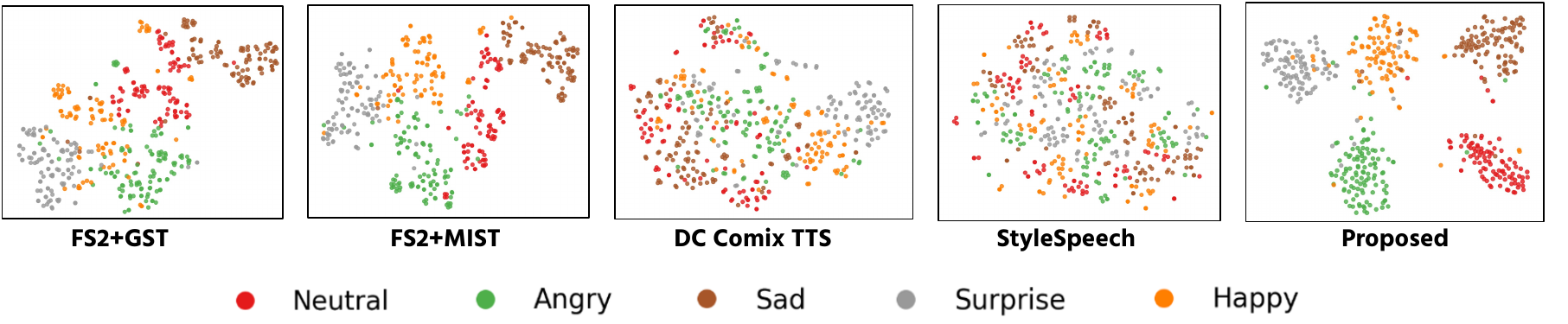}
    \caption{\normalfont T-SNE visualization of emotion embeddings}
    \label{tsne}
\end{figure*}

We conducted both objective and subjective evaluations to comprehensively assess our system.

\noindent\textbf{Subjective metrics.} Naturalness was evaluated using the classical Mean Opinion Score (MOS), where ten judges rated each utterance on a 1-to-5 scale (1 = bad, 2 = poor, 3 = fair, 4 = good, 5 = excellent). To assess emotional or style similarity between reference and synthesized speech, we adopted the Similarity MOS (SMOS), in which the judges scored each utterance pair on a 4-point scale: 1 = Very dissimilar, 2 = Dissimilar, 3 = Similar, 4 = Very similar. All MOS and SMOS results were reported together with their 95\% confidence intervals (CI).

\noindent\textbf{Objective metrics.} Spectral fidelity was measured by mel-cepstral distortion (MCD), where synthetic and reference mel-spectrograms were first aligned via dynamic time warping (DTW). Expressive adequacy was gauged by an emotion-recognition task: we fine-tuned \texttt{openai/whisper-large-v2}\footnote{\url{https://huggingface.co/openai/whisper-large-v2}} on the ESD training split and report the resulting unweighted average accuracy (UAA) on generated speech.

\subsection{Results}



Across all evaluations we prevented ``content leakage''—the artificial boost that arose when the reference speech shared lexical content with the synthesis target.  
Specifically, each reference mel-spectrogram was drawn (with a fixed random seed) from an utterance that matched the speaker and emotion of the target but \emph{differed in text}.  
Table~\ref{tab:results1} \textit{reported} MOS, MCD, and UAA obtained with two pre-trained recognizers.  
Our model attained naturalness that is statistically on par with the best baseline, while delivering markedly superior style consistency, underscoring its effectiveness at reproducing the intended style without sacrificing perceptual quality.

To visualize the extent of feature disentanglement, Figure~\ref{tsne} \textit{showed} a t-SNE projection of the emotion embeddings.  
Because our Emotion Extractor \textit{generated} phoneme‐level emotion embeddings, we \textit{averaged} them to a single utterance‐level embedding before projection.  
The proposed method \textit{yielded} tight, well-separated clusters for the five emotion categories, whereas embeddings from the strongest baseline \textit{scattered} widely and \textit{overlapped} across classes.  
This qualitative evidence \textit{corroborated} the quantitative gains and \textit{highlighted} the efficacy of our disentanglement strategy.

\subsection{Ablation Study}

\begin{table}[t]\small
\centering
\footnotesize
\caption{\normalfont Results of ablation study (emotion/speaker predictors \& MINE)}
\begin{tabular}{@{}lccccc@{}}
\toprule
\textbf{Model} & \textbf{MOS(↑)} & \textbf{SMOS(↑)}    & \textbf{MCD(↓)}   & \textbf{UAA(↑)}  \\ \midrule
\textbf{Proposed}  & $\mathbf{3.62\pm0.10}$   & $\mathbf{3.54\pm0.06}$    & $\mathbf{8.23\pm0.22}$ & $\mathbf{82.42\%}$\\ \midrule
w/o Predictors   & $3.53\pm0.10$    & $3.29\pm0.07$       & $9.71\pm0.21$      & 56.84\%         \\
w/o MINE    & $3.50\pm0.09$  & $3.47\pm0.06$    & $8.59\pm0.22$      & 76.37\%    \\ \bottomrule
\end{tabular}
\label{tab:results2}
\end{table}

To validate the effectiveness of individual components in our proposed model, we conducted ablation experiments, and the results were presented in Table~\ref{tab:results2}. In this table, ``w/o Predictors'' denoted the removal of the Emotion and Speaker Predictors, and their corresponding loss functions were not optimized, whereas ``w/o MINE'' denoted the exclusion of MINE, meaning the mutual information between the outputs of the timbre and emotion extractors was not minimized. The experimental results demonstrated that incorporating both MINE and the Emotion and Speaker Predictors significantly improved performance. Specifically, the combined use of these components allowed the extractors to better distinguish and capture distinct features from the reference speech. This, in turn, enhanced the overall quality of the synthesized speech.

\section{Conclusions}
The study introduced a novel emotional TTS method, enhancing the FS2 architecture with a phoneme-level Emotion Extractor and global Timbre Extractor. To achieve effective disentanglement of style representations, we leveraged a MINE to minimize the mutual information between different feature dimensions. Experimental results demonstrated that our approach consistently outperformed baseline models, highlighting its efficacy.

For future work, we plan to extend our phoneme-level emotion embedding and style disentanglement techniques to multimodal generation and conversational speech dialogue systems, where fine-grained controllability and robust style transfer are equally crucial.

Moreover, we acknowledge that our current backbone is FastSpeech2, which is not state-of-the-art in naturalness and expressivity. As future work, we will port our phoneme-level emotion embedding and disentanglement to diffusion-based and language-model–based TTS backbones.

\textbf{Acknowledgment:}
The work was supported by JSPS KAKENHI Grant Numbers 25K21266 and 22K12069, JST Moonshot Grant Number JPMJMS2011 and based on results obtained from a project outsourced by the New Energy and Industrial Technology Development Organization (NEDO).

\printbibliography

\end{document}